%% file: main.tex
\documentclass[sigconf]{acmart}

\AtBeginDocument{%
  \providecommand\BibTeX{{%
    \normalfont B\kern-0.5em{\scshape i\kern-0.25em b}\kern-0.8em\TeX}}}

\copyrightyear{2024}
\acmYear{2024}
\setcopyright{acmlicensed}\acmConference[KDD '24]{Proceedings of the 30th
ACM SIGKDD Conference on Knowledge Discovery and Data Mining}{August
25--29, 2024}{Barcelona, Spain}
\acmBooktitle{Proceedings of the 30th ACM SIGKDD Conference on Knowledge
Discovery and Data Mining (KDD '24), August 25--29, 2024, Barcelona, Spain}
\acmDOI{10.1145/3637528.3671513}
\acmISBN{979-8-4007-0490-1/24/08}

\usepackage[utf8]{inputenc}

\usepackage{amsmath,amssymb,amsfonts}
\usepackage{algorithmic}
\usepackage{graphicx}
\usepackage{textcomp}
\usepackage{xcolor}

\usepackage{siunitx}
\usepackage{multirow}
\usepackage{array}
\usepackage{booktabs}
\usepackage[most]{tcolorbox}
\usepackage{lineno}
\usepackage{balance}

\newcommand{\ie}{\textit{i.e.}}
\newcommand{\eg}{\textit{e.g.}}

\newcommand{\ours}{{CAMA}}
\newcommand{\huawei}{{Huawei Cloud}}

\newcommand{\todo}[1]{\{\textcolor{blue}{\textbf{TODO}}\}}

\newtcolorbox[list inside=prompt]{prompt}[1][]{
    colbacktitle=black!60,
    coltitle=white,
    fontupper=\footnotesize,
    boxsep=5pt,
    left=0pt,
    right=-1pt,
    top=0pt,
    bottom=0pt,
    boxrule=1pt,
    #1,
}

\begin{document}

% \title{Multi-field Matching for Cloud Solution}
\title{Enhancing Multi-field B2B Cloud Solution Matching via Contrastive Pre-training}

\author{Haonan Chen}
\affiliation{%
  % \institution{Gaoling School of Artificial Intelligence, Renmin University of China}
  \institution{Renmin University of China}
  \city{Beijing}
  \country{China}}
\email{hnchen@ruc.edu.cn}

\author{Zhicheng Dou}
\affiliation{%
  % \institution{Gaoling School of Artificial Intelligence, Renmin University of China}
  \institution{Renmin University of China}
  \city{Beijing}
  \country{China}}
\email{dou@ruc.edu.cn}

\author{Xuetong Hao}
\affiliation{%
  % \institution{Algorithm Innovation Lab, Huawei Cloud Computing Technologies}
  \institution{Huawei Cloud Computing}
  \city{Hangzhou}
  \country{China}}
\email{haoxuetong@huawei.com}

\author{Yunhao Tao}
\affiliation{%
  % \institution{Gaoling School of Artificial Intelligence, Renmin University of China}
  \institution{Renmin University of China}
  \city{Beijing}
  \country{China}}
\email{yunhao@ruc.edu.cn}

\author{Shiren Song}
\affiliation{%
  % \institution{Gaoling School of Artificial Intelligence, Renmin University of China}
  \institution{Renmin University of China}
  \city{Beijing}
  \country{China}}
\email{shiren.song@ruc.edu.cn}

\author{Zhenli Sheng}
\affiliation{%
  % \institution{Algorithm Innovation Lab, Huawei Cloud Computing Technologies}
  \institution{Huawei Cloud Computing}
  \city{Hangzhou}
  \country{China}}
\email{shengzhenli@huawei.com}

\begin{abstract}

Cloud solutions have gained significant popularity in the technology industry as they offer a combination of services and tools to tackle specific problems. 
However, despite their widespread use, the task of identifying appropriate company customers for a specific target solution to the sales team of a solution provider remains a complex business problem that existing matching systems have yet to adequately address.
In this work, we study the B2B solution matching problem and identify two main challenges of this scenario: (1) the modeling of complex multi-field features and (2) the limited, incomplete, and sparse transaction data.
To tackle these challenges, we propose a framework \ours{}, which is built with a hierarchical multi-field matching structure as its backbone and supplemented by three data augmentation strategies and a contrastive pre-training objective to compensate for the imperfections in the available data. 
Through extensive experiments on a real-world dataset, we demonstrate that \ours{} outperforms several strong baseline matching models significantly. 
Furthermore, we have deployed our matching framework on a system of \huawei{}. 
\textbf{Our observations indicate an improvement of about 30\% compared to the previous online model in terms of Conversion Rate (CVR), which demonstrates its great business value.}

\end{abstract}




\maketitle

\section{Introduction}

Cloud solutions, referring to a combination of various cloud-based technologies, tools, and services, have become increasingly popular among companies these years. 
They are designed to address specific business needs or solve particular problems, such as data storage, application development, customer relationship management (CRM), etc. 
For solution providers, it is crucial to have an effective matching system that can guide sales teams in identifying potential enterprises that can buy the solution. 
This is particularly important due to the marketing value of solutions and the high cost of human resources in Business-to-Business (B2B) scenarios.

While there have been some studies focusing on designing effective matching systems~\cite{multiranking2018, multidr2018, multids2022, multidr2023, ditto, DeepMatcher, HierGAT}, none of these works have explored the matching of cloud solutions and their customers, which holds significant business value.
In \huawei{}, the scenario is manual-driven, wherein our model identifies a list of the top matching companies to the sales team associated with a specific solution. 
The sales team then manually reviews this list and proceeds with promoting the solution to those companies.
This specific scenario can be considered a matching problem, with the primary goal being the identification of appropriate companies (customers) for the sales teams to target in their promotion efforts.

In this work, we focus on this specific scenario of B2B solution matching and identify \textbf{two main challenges}:
\textbf{(1) The features of solutions and companies are complex and often comprised of multiple fields.}
As presented in Table~\ref{tab:example}, the features consist of text, categorical, and numeric features, which consist of multiple fields.
Modeling these different types of features can pose challenges, such as different encoding paradigms for texts and other features, potential interference between different text fields, and interactions between different types of features.
\textbf{(2) The available transaction data, which include recorded successful purchases, are limited, incomplete, and sparse.}
The paradigm of our scenario is manual-driven, \ie, the list of highly matched companies generated by a matching system needs to be manually reviewed by sales teams before contacting potential customers.
As a result, the matching of solutions and target companies in a B2B scenario requires significant human resources, leading to limited recorded data.
Additionally, the data of solutions and companies may be incomplete, with specific fields or tokens missing due to recording or registration errors (as shown in Table~\ref{tab:example}).
In addition, numerous potential purchases remain undiscovered. 
For instance, despite similar companies requiring the same solution, they may not be persuaded by the sales team due to various factors, including the effectiveness of the sales pitch, the company's history of purchasing solutions from other providers, and even personal relationships. 
% These subjective factors can not be measured by a matching system.
Consequently, our available training data inherently suffers from data sparsity.
% \textbf{(3) There is a high demand for interpretability.}
% Given the high cost of B2B recommendations, our recommender system needs to provide detailed and evidential reasons for recommendations to sales teams, enabling them to pursue sales opportunities effectively.

To address these problems, we propose a \textbf{C}ontrastive pre-trained hier\textbf{A}rchical \textbf{M}ulti-field m\textbf{A}tching framework for B2B cloud solution matching (\ours{}). More specifically:

\noindent $\bullet$ \textbf{For the first challenge}, we propose a hierarchical multi-field matching framework as the backbone of \ours{}. 
% This framework allows us to effectively model the interactions between solutions and companies with multiple complex fields.
To mitigate token-level interference between different types of texts, we separate the text fields into description-like texts and attribute-like texts. 
% The reasons for this separation are discussed in detail in Section~\ref{subsec:problem definition}.
To gain a comprehensive understanding of each text group, we utilize two separate BERT models~\cite{bert} as token-level encoders. 
% This enables us to capture the nuances and context of the texts more accurately. 
We further incorporate Field-aware Embeddings into the embedding layers of BERTs to identify different fields during the matching process.
In addition to the text fields, we also employ look-up embeddings and the AutoDis technique~\cite{AutoDis} to encode categorical and numerical features, respectively.
Furthermore, we model the interactions among different feature groups at a higher level using another Transformer encoder~\cite{transformer}. 
This allows us to capture the dependencies and relationships between the various feature groups.

\noindent $\bullet$ \textbf{To address the second challenge}, we devise several data augmentation strategies and implement a contrastive learning objective to pre-train the BERT encoders.
To generate additional solution-company sample pairs, we employ three augmentation techniques: Token Masking, Field Masking, and Company Replacing. 
These strategies allow us to complement the imperfect transaction data by introducing variations.
Through pre-training the BERT encoders with augmented data and a contrastive learning objective, we aim to enhance our model's robustness and generalization. 
% This approach enables us to capture the intricate relationships and dependencies between the texts, even with imperfect data.
% \textbf{To deal with the last challenge}, we devise a two-step interpretation module.
% Firstly, we extract valuable information from the complex input features by attributing each matching score to its corresponding input. 
% % This allows us to understand the contributions of different features and identify the most relevant ones for the matching task.
% Secondly, we leverage a Large Language Model (LLM) to analyze and summarize the extracted features into a readable explanation for the sales team. 
% This summary provides a concise analysis of the key factors that contribute to the scores, enabling the sales team to understand the reasoning behind the recommendations.

\begin{table}[t]
    \centering
    \small
    \caption{An example solution-company pair to present their complex multi-field features. The missing fields and tokens due to recording errors are marked with the color \textcolor{red}{red}.}
    \begin{tabular}{lll}
    \toprule
        Group & Field & {Feature}  \\
        \midrule
        \multicolumn{3}{c}{Solution}  \\
        \midrule
        \multirow{2}{*}{\begin{tabular}[c]{@{}c@{}}Description \\ ($s^\text{d}$)\end{tabular}} 
        & Name & Distributed Cache Service  \\
        
        & Introduction  & There are currently hot data in... \\ 
        \midrule
        \multirow{4}{*}{\begin{tabular}[c]{@{}c@{}}Attribute \\ ($s^\text{a})$\end{tabular}} 
        & First-level Industry & Internet   \\
        
        & Second-level Industry & Internet Information Services  \\ 
        
        & Third-level Industry & \textcolor{red}{NA}  \\ 
        
        & Target Industry & \textcolor{red}{NA}  \\ 
        
        \midrule
        \multicolumn{3}{c}{Company}  \\
        \midrule
        \multirow{3}{*}{\begin{tabular}[c]{@{}c@{}}Description \\ ($c^\text{d}$)\end{tabular}} 
        & Name & *** (masked for privacy concern)   \\
        
        & {Introduction}  & *** is an internet education... \\
        
        & {Business Scope} & The business scope includes... \\ 
        \midrule
        \multirow{5}{*}{\begin{tabular}[c]{@{}c@{}}Attribute \\ ($c^\text{a}$)\end{tabular}} 
        
        & First-level Industry & Education  \\
        
        & Second-level Industry & \textcolor{red}{(Online)} Education  \\ 
        
        & Third-level Industry & {Skills training}, ...  \\ 
        
        & Copyright Name & review system, ...  \\ 
        
        & Key Project Category & Education Platform \\ 
        
        \midrule
        \multirow{2}{*}{\begin{tabular}[c]{@{}c@{}}Scale \\ ($c^\text{s}$)\end{tabular}} 
        & Categorical Features & \{Status: 1\}, ... (8 more features)   \\
        
        & Numeric Features & \{\# App: 7\}, ... (21 more features) \\

    \bottomrule
    \end{tabular}
    \label{tab:example}
\end{table}

Our experiments in both online and offline settings demonstrate the effectiveness of our proposed model.
\ours{} performs significantly better than some strong matching baseline models on a real-world dataset.
After being deployed on an online system, it surpasses the previous online model of \huawei by about 30\% in terms of Conversion Rate.

In summary, the contributions of our work are as follows:

(1) We recognize the value and importance of the B2B cloud solution matching problem and identify two major challenges.

(2) To address these challenges, we propose \ours{}, a framework that incorporates a hierarchical multi-field matching approach and a text-matching enhancement module utilizing contrastive learning.

(3) We validate the effectiveness of our framework through experiments conducted in both offline and online settings. 
The results demonstrate that our method is effective in the matching of cloud solutions and target companies in real-world B2B scenarios, offering substantial industrial value.

\section{Related Work}
% \todo{more content?}
\subsection{B2B Application Scenario}

Business-to-Business (B2B) systems differ from Business-to-Consumer systems typically employed on e-commerce platforms. 
B2B systems are designed for more complex scenarios and are primarily deployed within a company for internal usage. 
Consequently, the existing literature on B2B matching is limited.
Zhang et al.~\cite{zhang2005study} was an early work introducing B2B matching scenario.
Some works combined several techniques to design hybrid systems~\cite{hybird2015,hybird2018, hybird2021}.
For instance, Pande et al.~\cite{hybird2021} combined case-based reasoning (CBR), Collaborative Filtering (CF), and Random Walk for Consultancy.
Furthermore, some applied tree-based algorithm~\cite{treebased2015} and graph-based methods~\cite{GCN2021, graph2015, graph2018} to model relationships of customers.
For example, Henna et al.~\cite{GCN2021} utilized a graph convolution network (GCN) for B2B customer relationship management.
% Unlike B2C systems, a B2B system needs detailed and evidential recommendation interpretations for salesmen to pursue sales opportunities.
% Additionally, some works attempted to design interpretable B2B recommender systems~\cite{interpretable2017}
These works all make great contributions to this field. 
However, none of these studies have explored the valuable solution-company matching scenario and its associated challenges.

\begin{figure*}[t!]
    \centering
    \includegraphics[width=.9\linewidth]{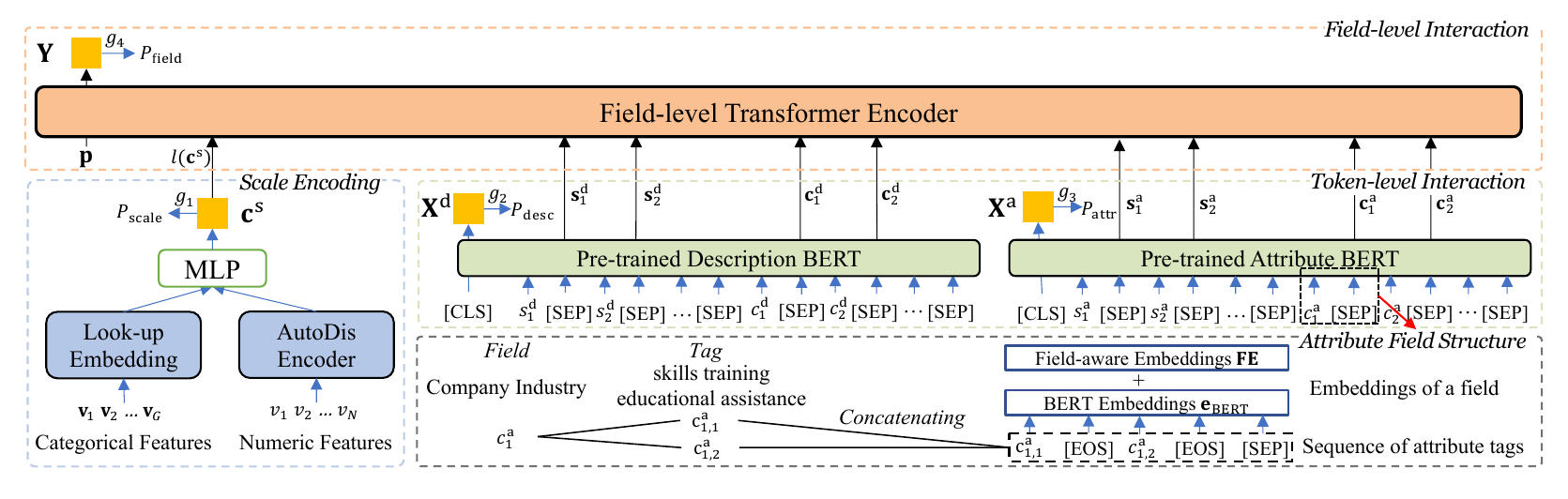}
    \caption{The illustration of \ours{}. The scale encoding module incorporates the usage of look-up embedding and the AutoDis encoder to effectively model categorical and numeric features, respectively. Furthermore, two pre-trained BERT encoders are employed along with field-aware embeddings to capture token-level interactions within two distinct groups of text pairs. At a higher level, a Transformer encoder is utilized to model field-level interactions among various feature groups.}
    \label{fig:framework}
    \vspace{-5px}
\end{figure*}

\subsection{Matching Models}

In our scenario, the problem can be considered a matching problem of solutions and companies, \ie, helping salesmen contact the companies that match the solution.
In neural text matching, researchers focused on two kinds of models, representation-based~\cite{dssm, sbert, twot2015, twot2016, twot2020, twot2021, colbert2020, convaug, LLM4IRSurvey} and interaction-based~\cite{onet2016, onet2017, bert, onet2021, graphformer, ricr, onet2023}.
Representation-based models convert sentences into hidden vectors, whereas interaction-based models match texts on word-level interactions. 
Because of the high cost of human marketing, there is a high requirement for B2B matching systems' accuracy.
Thus, we need to model the interactions of the texts.
% To model solutions and companies' complex fields of various types, our model designs a multi-field matching framework.

Some researchers have studied multi-aspect text matching for News Recommendation~\cite{multirs2015, multinews2019, multinews2021}, Document Ranking~\cite{multiranking2018}, and Dense Retrieval~\cite{multidr2018, multids2022, multidr2023, llmcs}.
For instance, Kong et al.~\cite{multids2022} represented multiple aspects of a query using different embeddings.
Shan et al.~\cite{multidr2023} designed an attribute-guided representation learning framework to couple the query and item representation learning together. 
This framework can also identify the most relevant item features for item representation.
To facilitate text matching, we also encode the scale features of companies.
Some works have studied to incorporate side information into the text-matching process for Document Ranking~\cite{multirank2020} and Recommender Systems~\cite{FM, DeepFM, wide&deep}.
Wide\&Deep~\cite{wide&deep} combined a wide linear model and a DNN to capture both sparse features and dense embeddings.
Some works of Entity Matching have developed matching frameworks that can be potentially applied to our cloud solution matching problem~\cite{DeepMatcher, ditto, HierGAT}.
For example, HierGAT~\cite{HierGAT} developed a hierarchical graph attention transformer that utilizes both self-attention and graph-attention mechanisms.

\section{Methodology} \label{sec:method}

This study focuses on the matching of solutions and companies in the B2B scenario, which holds significant commercial value but has received limited attention in previous research. 
% We identify three key challenges associated with this scenario. 
To address the challenges of this scenario, we first propose a Hierarchical Multi-field Matching framework to model the complex multi-field features of solutions and companies. 
% categorize these features into three groups: description texts, attribute texts, and scale features. 
Specifically, by considering three aspects of the modeling process, \ie, scale features encoding, fine-grained token-level interaction, and field-level inter-group interaction, we can compute matching scores from different perspectives.
Furthermore, recognizing the issue of limited, incomplete, and sparse transaction data, we devise several data augmentation strategies to generate supplemental solution-company data pairs. 
Additionally, we employ a contrastive learning objective to pre-train our text models, enhancing their ability to learn the intricate interactions between solutions and companies.
% To meet the interpretability requirement of B2B recommendation systems, we further design an interpretability module. 
% This module identifies crucial texts and scale features by attributing the matching scores to corresponding inputs. 
% Subsequently, it employs an LLM to summarize these extracted features into a coherent and comprehensible explanation, thereby facilitating the understanding of the recommendations by sales teams.

\subsection{Problem Definition} \label{subsec:problem definition}

Before shedding light on our model, we first give a concise definition of the problem we study.
Our objective is to identify potential buyers (companies) to the sales teams based on the outcomes of our framework.
Specifically, we denote the solutions as $S$ and the companies as $C$.
For each solution $s \in S$, we need to rank $C$ based on the matching scores between $s$ and every company $c \in C$, denoted as $P(s,c)$.
% To simplify our discussion in this paper, we will focus on the matching of one solution $s$ and one company $c$.
As presented in Table~\ref{tab:example}, the fields of $s$ and $c$ are divided into three groups: description texts ($s^\text{d}$, $c^\text{d}$), attribute texts ($s^\text{a}$, $c^\text{a}$), and scale features ($c^\text{s}$) (only $c$ has categorical and numerical features representing its scale).
We categorize the text features into two distinct groups based on their meanings and structures:
(1) The description texts are typically long natural language texts consisting of general descriptions of $s$ and $c$.
On the other hand, attribute texts include keywords or tags that represent attributes of $s$ and $c$.
Due to the distinct nature of these two types of texts, their token-level interactions can interfere with each other (demonstrated in Section~\ref{subsec: ablation}). 
% Therefore, it is necessary to separate them into different groups.
(2) The text features of solutions and companies are heterogeneous.
In other words, the features of $s$ and $c$ are not exact matches.
Thus, it is not feasible to treat each field as a group and perform field-to-field matching between $s$ and $c$.
% To address these problems, we opt to categorize the text features into two groups based on their meanings and structures.

The top-ranked company list of $C$ will be distributed to the sales team responsible for promoting the corresponding solution, who will subsequently contact these companies and pursue potential sales opportunities. 
% The primary objective of this matching task is to develop a comprehensive model capable of effectively capturing the intricate multi-field features inherent in both $s$ and $c$. 
By leveraging the multi-field features, the model aims to learn the patterns of matching between $s$ and $c$ and subsequently rank the companies most likely to purchase $s$ as high as possible within the generated lists. 
% The ultimate goal is to optimize the ranking process and maximize the potential for successful sales engagements.

\subsection{Framework Overview} \label{subsec:overview}

In this part, we will briefly introduce the overall structure of our framework.
Our framework is comprised of two parts:

(1) \textbf{Hierarchical Multi-field Matching.}
As shown in the lower part of Fig.~\ref{fig:framework}, we design a hierarchical matching structure to effectively capture the interactions between solutions and companies whose features are comprised of complex fields.
To begin with, we focus on capturing the fine-grained token-level interactions within two groups of text fields. 
This is achieved by utilizing BERT encoders and field-aware embeddings, which allows us to extract rich representations from the textual data.
We also encode scale features of $c$ into a representation that captures the company's scale.
Subsequently, we employ a Transformer encoder to model the field-level inter-group interactions. 
% This helps us capture the dependencies between different fields in a comprehensive manner.
Four matching scores are calculated based on the interactions from different perspectives.
% In addition, the representation of $c^\text{s}$ is used to complement the field-level understanding of the texts.
% Finally, the output representations from the token-level encoders, scale encoder, and field-level encoder are ensembled by a classifier into the matching score of ($s$, $c$).

(2) \textbf{Text Matching Enhancement.}
In this part, we attempt to enhance the token-level interactions of the text features.
To achieve this, we design three data augmentation strategies and a contrastive objective to pre-train the BERT encoders.
Specifically, for each $(s,c)$ pair in the training data, we randomly select two strategies to generate two similar pairs.
% To model the interactions between these pairs, we utilize a BERT encoder and consider the output representation of the special token ``[CLS]'' as the representation of the interaction. 
Subsequently, we apply a contrastive loss function to pull together the representations of the generated pairs and push them away from other pairs within the same mini-batch.
By pre-training our BERT encoders with this objective, we effectively improve the modeling of text interaction, especially with limited, incomplete, and sparse data.

% (3) \textbf{Two-step Interpretation.}
% As shown in the higher part of Fig.~\ref{fig:framework}, we propose a two-step approach to generate a comprehensive and easily understandable explanation that includes important information about the input features. 
% Firstly, we employ Integrated Gradients~\cite{IntegratedGradients} to attribute the predictions of token-level BERT models to their respective input tokens, allowing us to extract valuable phrases. 
% We also attribute the prediction of the field-level Transformers to identify important fields. 
% In the case of scale features, we remove each feature individually and evaluate its impact on the model's performance. 
% Once we have extracted these important features, we employ an LLM to analyze and summarize them into a concise and reader-friendly explanation tailored for sales teams.

\subsection{Hierarchical Multi-field Matching} \label{subsec:HMM}

In our B2B cloud solution matching scenario, we have identified a significant and unexplored challenge: the modeling of complex multi-field feature interactions.
Specifically, the fields of solutions and companies are comprised of two main kinds of features: scale features and text features.
Consequently, we devise distinct models to effectively capture and analyze these different types of features.

\subsubsection{Scale Encoding}

Since only $c$ has scale features that represent its scale, our goal here is to encode these features into a representation rather than modeling interactions.
Instead of using $c^\text{s}$ directly, we encode it to better interact with the textual representations.
Besides, $c^\text{s}$ is comprised of both categorical (\eg, whether $c$ is listed) and numerical (\eg, registration capital) fields.
Therefore, we encode these two types of features separately and fuse them into a unified representation, as illustrated in the lower left part of Fig.~\ref{fig:framework}.

Suppose $c^\text{s}$ contains $G$ categorical fields and $N$ numerical fields: $c^\text{s} = [\mathbf{v}_1, \mathbf{v}_2, \ldots, \mathbf{v}_{G}; {v}_1, {v}_2, \ldots, {v}_N]$, 
where $\mathbf{v}_i$ is the one-hot vector of the value of the $i$-th categorical field, and $v_j$ is the scalar value of the $j$-th field of the numerical features.

For the categorical fields, we apply a look-up embedding technique to encode the one-hot vectors.
Specifically, for the $i$-th categorical field, we obtain its embedding: $\mathbf{e}_i = \mathbf{E}_i \cdot \mathbf{v}_i,$
where $\mathbf{E}_i \in \mathbb{R}^{f_i \times d_\text{s}}$ is the embedding matrix for the $i$-th field to look-up, $f_i$ is the field size, and $d_\text{s}$ is the embedding size of all scale features. 

To handle the numerical fields, we employ an automatic embedding learning technique based on soft discretization (AutoDis~\cite{AutoDis}).
The utilization of soft discretization within our end-to-end learning framework allows for the optimization of this process.
% Unlike hard discretization methods, AutoDis has the advantage of encoding each feature into a distinct representation, resulting in superior performance.
% Therefore, we apply the AutoDis technique to encode our numerical features.
First, the scalar value $v_j$ is discretized into buckets by a two-layer neural network with skip-connection $\mathbf{h}_j = {\rm Leaky\_ReLU}(\mathbf{w}_j v_j), \widetilde{\mathbf{v}}_j = \mathbf{W}_j \mathbf{h}_j + \alpha \mathbf{h}_j,$
where $\mathbf{w}_j \in \mathbb{R}^{1 \times H_j}$ and $\mathbf{W}_j \in \mathbb{R}^{H_j \times H_j}$ are learnable parameters that automatically discretized $v_j$ into the projection outputs of $H_j$ buckets: $\widetilde{\mathbf{v}}_j = [\widetilde{v}_j^1, \widetilde{v}_j^2, \ldots, \widetilde{v}_j^{H_j}]$, where $\widetilde{v}_j^h$ is the projection of scalar value $v_j$ on the $h$-th bucket.
This projection is then normalized by the Softmax(·) function into a weight on the corresponding bucket: $\widehat{v}_j^h = {\rm Softmax}(\widetilde{v}_j^h)$.

Subsequently, a set of meta-embeddings $\mathbf{ME}_j \in \mathbb{R}^{H_j \times d_\text{s}}$ is designed for the $j$-th field.
The softly discretized results $\widehat{\mathbf{v}}_j$ represents the relevance between the $j$-th field of the numeric features and the buckets of meta-embeddings.
Thus, we can leverage a weighted-average technique to aggregate the meta-embeddings and their corresponding weights into a representation for each numerical feature: $\mathbf{e}^{\text{num}}_j = \sum_{h=1}^{H_j}\widehat{v}_j^h \cdot \mathbf{ME}_j^h$.

After the scale features are embedded into continuous vectors, we employ a Multi-Layer Perceptron (MLP) to fuse them into a unified representation $\mathbf{c}^\text{s} \in \mathbb{R}^{d_\text{s}}$ that captures the scale features of $c$: $\mathbf{c}^\text{s} = {\rm MLP} \big([\mathbf{e}_1, \mathbf{e}_2, \ldots, \mathbf{e}_{G}; \mathbf{e}^{\text{num}}_1, \mathbf{e}^{\text{num}}_2, \ldots, \mathbf{e}^{\text{num}}_{N}]\big).$
We can get a score by applying a linear projection $g_1(\cdot)$ to map this representation into a (scalar) score: $P_{\text{scale}}(s,c) = g_1(\mathbf{c}^\text{s})$.
\subsubsection{Token-level Interaction} \label{subsubsec:token match}
In this part, we attempt to model the token-level interactions of $(s,c)$.
Pre-trained language models, such as BERT~\cite{bert}, have gained significant popularity in various tasks, including Recommender Systems~\cite{BERT4Rec, RecBERT2022} and Information Retrieval~\cite{coca, ase}.
To capture the fine-grained token-level interactions of ($s^\text{d}$, $c^\text{d}$) and ($s^\text{a}$, $c^\text{a}$), we leverage BERT as the underlying encoder.
We employ special tokens to concatenate the fields in the description texts, resulting in the following sequence:
\begin{align}
    X^{\text{d}} &= {\rm [CLS]} s^\text{d}_1 {\rm [SEP]} \ldots s^\text{d}_{F_{\text{sd}}} {\rm  [SEP]} {\rm  [SEP]} c^\text{d}_1 \ldots c^\text{d}_{F_{\text{cd}}} {\rm  [SEP]}{\rm  [SEP]}, \notag
\end{align}
where $F$ is the number of text fields, $s_i$ and $c_j$ are the $i$-th and $j$-th fields that consist of many tokens of the solution and the company, respectively, ``${\rm [CLS]}$'' is the token used for representing the sequence, ``${\rm [SEP]}$'' is the separator token.
We append a ``${\rm [SEP]}$'' token after each field to indicate the end of a field and another one at the end of each sequence of fields.
Moreover, to distinguish the multiple tags in attribute texts (\eg, different industry tags), we further utilize ``${\rm [EOS]}$'' tokens to separate these tags before concatenating them (as shown in the lower center part of Fig.~\ref{fig:framework}):
\begin{align}
    c^\text{a}_j &= c^\text{a}_{j,1} {\rm [EOS]} c^\text{a}_{j,2} {\rm [EOS]} \ldots c^\text{a}_{j,T_j} {\rm [EOS]}, \notag
\\
% \end{align}
% \begin{align}
    X^\text{a} &= {\rm [CLS]} s^\text{a}_1 {\rm [SEP]} \ldots s^\text{a}_{F_{\text{sa}}} {\rm  [SEP]} {\rm  [SEP]} c^\text{a}_1 \ldots c^\text{a}_{F_{\text{ca}}} {\rm  [SEP]} {\rm [SEP]}, \notag
\end{align}
where $c^\text{a}_j$ is the $j$-th attribute field of the company and $T_j$ is the number of the tags contained in this field.

Furthermore, we design a set of field-aware embeddings that help the encoders to distinguish different fields during the modeling of interactions. 
Specifically, for each text group, we initialize a field-aware embedding matrix $\mathbf{FE} \in \mathbb{R}^{F \times d_\text{e}} $, where $F$ is the number of the fields and $d_\text{e}$ is the size of BERT's word embeddings.
The ``${\rm [EOS]}$'' tokens within and the ``${\rm [SEP]}$'' tokens after fields are also enhanced with corresponding field-aware embeddings (the $\mathbf{FE}$ of ``${\rm [CLS]}$'' is distinct from others).
Consequently, the embedding of each token is comprised of both our field-aware embedding and BERT embedding (as shown in the lower right part of Fig.~\ref{fig:framework}): $\mathbf{X}^\text{d} = {\rm BERT}^\text{d}_{{\rm [CLS]}}\big(\mathbf{FE}^\text{d} + \mathbf{e}_{\text{BERT}}(X^\text{d})\big), \mathbf{X}^\text{a} = {\rm BERT}^\text{a}_{{\rm [CLS]}}\big(\mathbf{FE}^\text{a} + \mathbf{e}_{\text{BERT}}(X^\text{a})\big),$
where $\mathbf{e}_{\text{BERT}}(\cdot)$ is the embedding layer of BERT.

We can get two token-level matching scores by applying linear projections $g_2(\cdot)$ and $g_3(\cdot)$ on $\mathbf{X}^{\text{d}}$ and $\mathbf{X}^{\text{a}}$, respectively:
    $P_{\text{desc}}(s,c) = g_2(\mathbf{X}^\text{d})$,
    $P_{\text{attr}}(s,c) = g_3(\mathbf{X}^\text{a})$.

\subsubsection{Field-level Interaction}
In the previous section, we have gathered information regarding token-level interactions within two distinct text groups.
However, although it has been observed that token-level interactions between different groups can have a detrimental effect on model performance, we still aim to capture the inter-group interactions from a higher level, \ie, the field level.
As shown in the center part of Fig.~\ref{fig:framework}, for each text field, we use the encoded output of the ``${\rm [SEP]}$'' token appended to it as its representation.
By doing so, we avoid using all tokens for the same reason we divide the features into two groups, \ie, preventing fine-grained interference.
Additionally, we use the encoded scale representation $\mathbf{c}^\text{s}$ to facilitate the modeling of field-level interactions.
This is because the scale of a company can influence how a solution interacts with it. 
For instance, interactions with smaller companies may place more emphasis on their copyright works due to their more focused business nature.

In order to comprehensively model these representations and capture the interactions among fields, we utilize the Transformer encoder, as proposed in the Transformer~\cite{transformer} architecture. 
The Transformer encoder effectively models the aforementioned representations in the following manner:
\begin{align}
    \mathbf{Y} = {\rm Trm}_{{\rm \mathbf{p}}} \big(\big[\mathbf{p}; {l}(\mathbf{c}^\text{s});~ &\mathbf{s}^\text{d}_{1}, \ldots, \mathbf{s}^\text{d}_{F_\text{sd}}, \mathbf{c}^\text{d}_{1}, \ldots, \mathbf{c}^\text{d}_{F_\text{cd}}; \notag \\
    &\mathbf{s}^\text{a}_{1}, \ldots, \mathbf{s}^\text{a}_{F_\text{sa}}, \mathbf{c}^\text{a}_{1}, \ldots, \mathbf{c}^\text{a}_{F_\text{ca}}\big]\big) ,
\end{align}
where ${\rm Trm}$(·) is the Transformer encoder which consists of $k$ Transformer layers, $\mathbf{s}$ and $\mathbf{c}$ are the encoded outputs of ``${\rm [SEP]}$'' appended to text fields, ${l}$(·) is a linear projector to map $\mathbf{c}^\text{s}$ into the latent space of the text field representations, and $\mathbf{p} \in \mathbb{R}^{d_\text{e}}$ is a randomly initialized vector used for pooling.

We can get a field-level matching score by applying a linear projection $g_4(\cdot)$:
$P_{\text{field}}(s,c) = g_4(\mathbf{Y})$.

% \subsubsection{Multi-grained Fusing}

% Up to this point, we have modeled the company scale, the token-level text interactions between $s$ and $c$, as well as the field-level interactions between them.
% To achieve a comprehensive outcome within our matching framework, we will leverage these encoded representations and employ a multi-grained fusing technique to calculate the matching score.
% The process is as follows:
% \begin{align}
%     P_{ensem}(s,c) = {\rm MLP} ([\mathbf{c}^\text{s}, \mathbf{X}^d, \mathbf{X}^a, \mathbf{Y}]),
% \end{align}
% where ${\rm MLP}$ is a multi-layer perception, and ${P_{ensem}(s,c)}$ is the (scalar) matching score.

\subsubsection{Optimization}
Since the label of whether a company buys a solution is binary (0/1), we use a cross-entropy loss to optimize our matching model.
We formulate the loss for the matching scores of our hierarchical models, \ie, $\{\mathcal{P}\} = \{P_{\text{scale}}, P_{\text{desc}}, P_{\text{attr}}, P_{\text{field}}\}$:
\begin{equation}
    \mathcal{L}_{\text{Match}} = -\frac{1}{M}\sum_{i=1}^M \sum\limits_{P \in \{\mathcal{P}\}}^{}  y_i\log{P_i}+(1-y_i)\log{(1-P_i)},
\end{equation}
where $M$ is the number of $(s,c)$ pairs in the training set, and $y_i$ is the label of the $i$-th data pair.

\subsection{Text Matching Enhancement}

\begin{figure}[t]
    \centering
    \includegraphics[width=\linewidth]{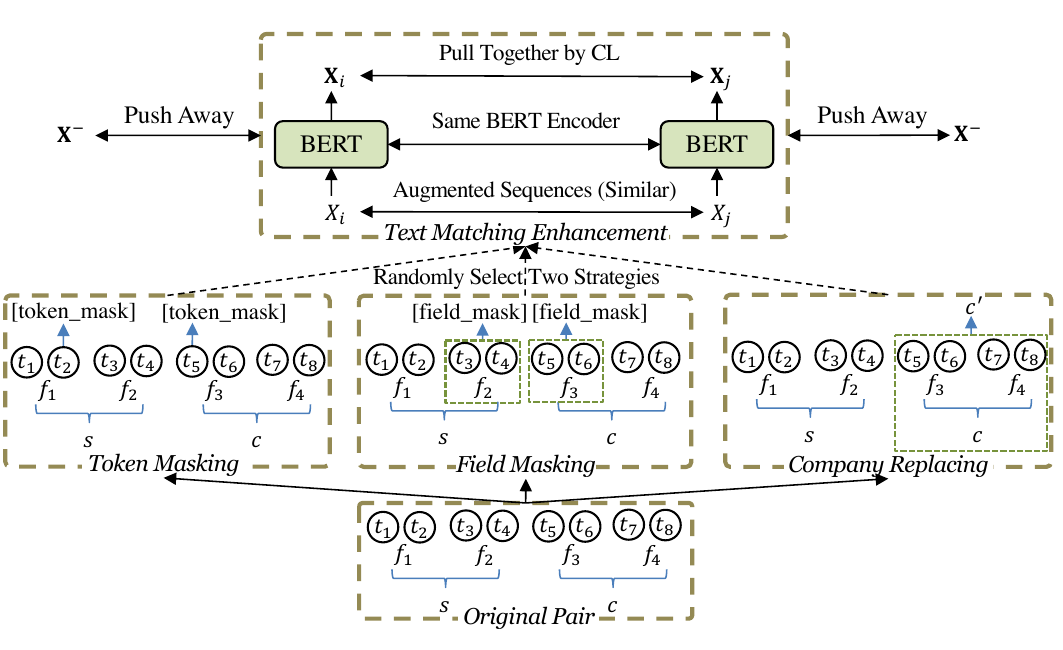}
    \caption{
    The illustration of our data augmentation strategies and contrastive learning process.
    Initially, an original $(s,c)$ pair is augmented by two random strategies. 
    The BERT-encoded representations of these two similar pairs are then brought closer through our contrastive loss function. 
    % Additionally, these representations are pushed away from other pairs within the same mini-batch.
    }
    \label{fig:data_aug}
\end{figure}

In our study, we identify several challenges of the transaction data in our scenario. 
Firstly, the training data is limited attributed to the high cost of human resources required for promoting solutions. 
Additionally, the data is incomplete as some solutions and companies lack specific tokens or fields. 
Moreover, the data is sparse, meaning that many potential pairs have not been discovered and recorded.
To mitigate these challenges, we attempt to enhance the generalization and robustness of the BERT encoders by pre-training them.
As shown in Fig.~\ref{fig:data_aug}, we employ a contrastive objective, which aims to bring together the representations of augmented similar sequences while pushing away different ones.

% Initially, an original (s,c) pair is augmented using two random strategies. Subsequently, the BERT-encoded representations of these augmented pairs are brought closer together through our contrastive loss function. Additionally, these representations are pushed away from other pairs within the same mini-batch to ensure better discrimination.

\subsubsection{Data Augmentation Strategies}

The similar sequences are generated from the inputs of BERT encoders ($X$).
We design three data augmentation strategies to generate additional data to complement the limited transaction data in our scenario.

(1) \textbf{Token Masking.}
Some existing works in Natural Language Processing~\cite{term1, term2, term3} have employed token-level augmentation techniques to enhance the robustness of sentence representations.
% By masking certain tokens within the input sequence and generating similar sequences that differ only in specific tokens, it becomes possible to contrast these sequences with others. 
This approach enables our BERT encoders to acquire more robust and generalized representations with incomplete data that may lack specific tokens, by reducing reliance on specific tokens.

To begin, we represent the token-level input sequence $X$ as a token sequence: $X = [t_1, t_2, \ldots, t_{W}]$, where $W$ denotes the total number of tokens.
Next, we randomly mask a proportion $r_\text{t}$ of the tokens in $X$: $T_\text{m}=[t'_1, t'_2, \ldots, t'_{M_\text{T}}]$, where $t'_i$ is the token to be masked and $M_\text{T} = \lfloor r_\text{t} * W \rfloor$.
For each token $t_i \in X$, if it is marked to be masked in $T_\text{m}$, we will replace it with a special token ``[token\_mask]'', which is similar to ``[MASK]'' in BERT~\cite{bert}.

(2) \textbf{Field Masking.}
To obtain robust representations of incomplete data pairs, our BERT encoder models should avoid relying on specific data fields while encoding the whole matching sequences.
By contrasting sequences that are augmented through the masking of certain fields with other sequences, we can facilitate the learning of text matching for incomplete data within our BERT models.

We represent $X$ as a sequence of data fields: $X = [f_1, f_2, \ldots, f_{F}]$, where $F$ is the total number of fields.
We randomly mask a ratio $r_\text{f}$ of the fields: $D_\text{m}=[f'_1, f'_2, \ldots, f'_{F_\text{T}}]$, where $f'_i$ represents the token to be masked and $F_\text{T} = \lfloor r_\text{f} * F \rfloor$.
For $f_i \in X$, if it is marked to be masked in $D_\text{m}$, we will replace it with a special token ``[field\_mask]''.

% (3) \textbf{Field Reordering}
% In contrast to some works in Natural Language Processing that require strictly ordered sequences~\cite{transformer}, the field sequences of solutions and companies are flexible.
% In other words, the order of fields in the matching sequence should not affect the modeling of interactions.
% However, our training data is insufficient to facilitate the learning of this important capability. 
% As a result, we have developed an augmentation strategy on field level, \ie, field reordering.

% Specifically, for the field sequence $X = [f_1, f_2, \ldots, f_{F}]$, we split it into solution fields and company fields: $X = [f^s_1, \ldots, f^s_{F_s}; f^c_1, \ldots, f^c_{F_c}]$.
% To maintain the matching structure of the $(s,c)$ pair, we perform the reordering process separately on these two groups of fields.
% Within each sequence, we will randomly select two fields and exchange their positions.
% %This re-ordering process will be performed $F_R = \lfloor r_r * F \rfloor$ times.

(3) \textbf{Company Replacing.}
% In our transaction data, many potentially positive pairs are neglected due to the high cost of human resources required to promote solutions.
% As a result, our data inherently suffer from the sparsity problem.
To address the data sparsity problem, we design a rule-based method to identify similar companies\footnote{We only find similar companies because the number of companies is much larger than the number of solutions, and solutions are often distinctive.} that can replace the company in the {positive} pairs.
% We only perform this augmentation strategy on positive pairs because our original data only include records of successful sales, and negative pairs are randomly sampled (as introduced in Section~\ref{subsec:dataset}).
% Directly replacing companies of these negative pairs will introduce much noise.
% By augmenting the dataset with pairs that have similar representations to the original ones, we aim to provide contrasting examples that can help our model effectively handle the sparse data.
Specifically, for each sequence $X$ of positive pairs $(s,c)$, we find one similar company $c'$ to generate a pair $(s,c')$.
% Shiren Song
% Since the company name is the identifier of a company and captures information about its industry and business nature, we do not consider other complex fields, in order to facilitate rapid computation and reduce the task's difficulty.
Because the company name serves as the primary identifier, capturing information about the industry and business nature, we only use the name to identify similar companies. This simplification aims to expedite computations and diminish the complexity of the rule.
Our rule-based method randomly selects one of the five most similar companies of company $c$ to be designated as company $c'$. The similarity between companies is obtained by calculating the semantic similarity of company names using the sentence-transformers package~\cite{sbert}.
% This simple yet highly effective approach can help to obtain a generalized representation of the model on company similarities even under sparse data.
% Shiren Song

\subsubsection{Contrastive Learning Objective}

Inspired by several studies in Information Retrieval~\cite{coca, cl1, cl2, cl3, cl4}, we adopt a contrastive learning approach to enhance our model for text matching.
In order to carry out contrastive learning for our task, we define a loss function specifically for the contrastive prediction task. 
This task involves identifying similar augmented pairs within the set $\{\mathcal{X}\}$, which is constructed by randomly augmenting the original sequences of a minibatch.
Suppose a minibatch contains $M$ sequences, we randomly employ two of our augmentation strategies to generate $\{\mathcal{X}\}$ comprising $2M$ sequences.
Among these sequences, the two sequences generated from the same solution-company pair are considered a similar (positive) pair, whereas the remaining $2(M-1)$ augmented ones serve as negative samples $\{\mathcal{X}\}^-$.
We formulate the contrastive learning loss for a positive pair $(\mathbf{X}_i,\mathbf{X}_j)$ as follows:
\begin{align}
 \phi(\cdot) &= \exp\left(\frac{{\rm Sim}(\cdot)}{\tau}\right),\\
    \mathcal{L}_{\text{CL}}(i,j) &= -\log\frac{\phi(\mathbf{X}_i,\mathbf{X}_j)}{\phi(\mathbf{X}_i,\mathbf{X}_j) + \sum\limits_{\mathbf{X}^- \in \{\mathcal{X}\}^-}^{}\phi(\mathbf{X}_i,\mathbf{X}^-)},
\end{align}
where ${\rm Sim}(\cdot)$ is a similarity function which is defined as cosine similarity in our study, and $\tau$ is a hyperparameter temperature.

\section{Experimental Setup}

\subsection{Dataset} \label{subsec:dataset}

% Shiren Song
To evaluate the effectiveness of our proposed method, we conducted extensive experiments on a real-world dataset.
We first sampled from the real-world data and constructed an offline dataset \textit{B2B Solution Matching} (BSM). 
BSM contains three parts: solution data, company data, and transaction data. 
The solution dataset stores the text information of 27 solutions, such as solution names, detailed descriptions, and tags of their industries. 
The company dataset provides detailed company profiles for 533,784 companies, including company names, company registration capital, and their business scopes, etc. 
% Table~\ref{tab:fields} presents the statistics of different field groups in solution and company data.
The transaction data are derived from the marketing feedback of the sales teams online.
Whether a company buys a solution is relevant to not only its suitability but also many subjective factors (\eg, the sales team’s pitch).
Consequently, unsuccessful purchases are often noise and disregarded.

We constructed the data format as a pairwise form (solution, company) along with their corresponding label from the abovementioned raw dataset. 
% The feature fields are organized into three groups, including description text, attribute text, and scale features. 
Then the constructed dataset is split into three parts, where 70\% is used for training, 10\% is used for validating and the remaining 20\% is for testing. 
For each positive sample, we randomly sampled 4 negative companies from the company dataset. 
% As for the testing data, it will take a lot of time to evaluate the model if we match each solution with each of the companies (around 14 million instances). 
% Thus for each company in the database (if this company is not in the training or validation set), we randomly selected a solution as a negative instance, with the sampling probability determined by the proportion of that solution in the transaction data (popular solutions need more testing data to ensure the model's performance on them). 
% By this, we can train and evaluate \ours{} on the scale of the whole company database without taking too much time.
The statistics of our dataset are presented in Table~\ref{tab:Datasets}.

\begin{table}[t]
\centering
\caption{Statistics of the BSM dataset.} \label{tab:Datasets}
\begin{tabular}{p{0.2\textwidth}ccc}
\toprule
\textbf{Datasets} & \textbf{Training} & \textbf{Validation} & \textbf{Testing}\\
\midrule
\# total samples & 48,515 & 6,930 & 13,860 \\
\# positive samples & 9,703 & 1,386 & 2,772 \\
\# negative samples & 38,812 & 5,544 & 11,088 \\
\bottomrule
\end{tabular}
\end{table}

\subsection{Evaluation Metrics}
We comprehensively evaluate our proposed method through both offline and online evaluations.

(1) \textbf{Offline Metrics.} We use Mean Average Precision (MAP), Area Under Curve (AUC), Precision at $k$ (P@$k$, $k={10,100, 500}$), and Recall at position $k$ (R@$k$, $k={10,100,500}$) as offline metrics. The results are calculated by averaging across different solutions.

(2) \textbf{Online Metrics.} \label{sec:online_metrics}
% We verify the performance of our proposed model with \textit{Intention Proportion (IP)} as online metrics. 
% IP is defined as \eqref{eq:metrics} below. 
% $\# Intentional$ denotes the number of companies with intention and $\# Recommended$ means the number of companies that the sales teams actually contact and recommend their solutions.
% \begin{align}
%     IP = \frac{\# Intentional}{\# Recommended}
%     \label{eq:metrics}
% \end{align}
We verify the performance of our proposed model with \textit{Conversion Rate (CVR)} as online metrics. 
CVR is defined as: $\text{CVR} = \#~\text{Purchase}/\#~\text{Promoted}$,
$\#~\text{Purchase}$ denotes the number of companies that purchase the solutions and $\#~\text{Promoted}$ means the number of companies that the sales teams market their solutions.
% \begin{align}
%     \text{CVR} = \frac{\#~\text{Purchase}}{\#~\text{Promoted}}
%     \label{eq:metrics}
% \end{align}
% Because of the large cost of human resources and the lengthy sales cycles in B2B recommendation scenarios, conducting online A/B tests is not feasible. 
% Furthermore, the relatively limited company transaction data makes it challenging to accumulate a sufficiently large dataset for valuable A/B tests.
To evaluate \ours{}'s effectiveness, we will compare its performance to a previous online model of Huawei. 
% The rules of this rule-based method have been meticulously crafted by seasoned sales professionals who possess a comprehensive understanding of the characteristics of the solutions and successful customers.

% Shiren Song

\begin{table*}[t]
    \centering
    \caption{Overall results on BSM.
    ``$\dag$'' denotes our model outperforms all baselines significantly in paired t-test at $p<0.01$ level (with Bonferroni correction).
    The best performance is in bold and the second-best performance is underlined.}
    \begin{tabular}{lccccccccccc}
\toprule
         & {SBERT} & {MADR} & {CBERT} & {DM} & {W\&D} & {HSCM} & {Ditto} & {MCN} & {HieGAT}                          & \ours{}         \\ \midrule
MAP     & 0.1953    & 0.2259   & 0.2539    & 0.4084 & 0.4178   & 0.4534   & 0.4711    & 0.5392  & \underline{0.5799} & \textbf{0.6810}$^\dag$ \\
AUC     & 0.7135    & 0.7238   & 0.7721    & 0.8093 & 0.8141   & 0.8136   & 0.8206    & 0.8451  & \underline{0.8480} & \textbf{0.8528}$^\dag$ \\ 
Rec@10  & 0.1809    & 0.2187   & 0.3283    & 0.3468 & 0.3579   & 0.3740   & 0.3887    & 0.4176  & \underline{0.4361} & \textbf{0.4579}$^\dag$ \\
Rec@100 & 0.2698    & 0.3685   & 0.3947    & 0.4782 & 0.4947   & 0.5396   & 0.5570    & 0.6295  & \underline{0.6727} & \textbf{0.7394}$^\dag$ \\
Rec@500 & 0.4156    & 0.5731   & 0.5581    & 0.7211 & 0.7372   & 0.7503   & 0.7773    & 0.8035  & \underline{0.8783} & \textbf{0.9335}$^\dag$ \\
Pre@10  & 0.3679    & 0.4084   & 0.4723    & 0.5021 & 0.5183   & 0.5420   & 0.5682    & 0.6168  & \underline{0.6492} & \textbf{0.6807}$^\dag$ \\
Pre@100 & 0.2561    & 0.2795   & 0.2810    & 0.2988 & 0.3019   & 0.3121   & 0.3344    & 0.3642  & \underline{0.3973} & \textbf{0.4266}$^\dag$ \\
Pre@500 & 0.0620    & 0.1068   & 0.1034    & 0.1529 & 0.1551   & 0.1634   & 0.1659    & 0.1740  & \underline{0.2291} & \textbf{0.2688}$^\dag$ \\ \bottomrule
\end{tabular}
    % \vspace{-5px}
    \label{tab:result}
\end{table*}

\subsection{Baseline Models}

% Yunhao Tao
We compare our \ours{} with two kinds of baselines: 

(1) \textbf{Text Matching Models} only use the interactions of texts to get the matching score.
% $\bullet$~{DSSM}~\cite{dssm} obtains the representations of texts by a deep neural network (DNN) and learns implicit semantics by maximizing the cosine similarity between the embeddings of text pairs. 
% Based on the two-tower structure, 
$\bullet$~{Sentence-BERT (SBERT)}~\cite{sbert} encodes the solution and company with two BERTs separately and takes the cosine similarity of these representations as the matching score.
% designs an objective function to concatenate two representations and the difference between them to fit the classifier. 
$\bullet$~{MADR}~\cite{multids2022} is also a representation-based model that designs aspect learning tasks to extract representations of different aspects from texts and then fuses them by calculating the weighted sum.
$\bullet$~{Concatenating-BERT (CBERT)} is an interaction-based model that models the token-level interaction of the solution and company without identifying different text fields. 
% In our scenario, the data lack labels of text aspects required in MADR.
% Thus, we use categorical features of company scale as aspect information to fit the fusion network, which may result in poor performance compared to other multi-aspect matching models.

(2) \textbf{Side-aware Matching Models} have frameworks that can encode side information (scale) along with the text interaction module to get the representations for matching.
$\bullet$~{Wide \& Deep (W\&D)}~\cite{wide&deep} combines a wide linear model and a DNN. 
For its implementation, we utilize a two-tower BERT model to get text representations and fit them into the framework with scale features.
% $\bullet$~{DeepFM}~\cite{DeepFM} consists of {FM}~\cite{FM} and a deep neural network (DNN).
% We fit BERT-encoded text representations into DNN to extract high-order features and scale features into FM.
% FM converts features into low-dimensional vectors and calculates the inner product of any two feature embeddings based on matrix factorization as an interaction.
$\bullet$~{MCN}~\cite{multirank2020} conducts an element-wise matching of solutions and companies, and fuses the matching output with scale features to learn interactive information. 
For the encoding modules in MCN, we use a two-tower BERT to get the textual representations and our scale encoding module to model scale features. 
% Then, we map text and scale features to the same dimensional space and calculate the interaction between the vectors by taking their dot product, as described in the article.
% We fit these side-aware frameworks with some strong encoding modules, resulting in several strong matching baselines.
We also apply some entity matching frameworks to our solution matching problem by treating our text and numeric fields as different entities.
$\bullet$~{DeepMatcher (DM)}~\cite{DeepMatcher} uses the GRU-RNN model to learn the attribute embeddings of entities, which are aggregated for matching. 
$\bullet$~{Ditto}~\cite{ditto} applies pre-trained Transformer-based language models and optimization techniques to perform the sequence-pair classification problem.
$\bullet$~{HierGAT}~\cite{HierGAT} combines Transformer attention with hierarchical graph attention to effectively learn entity embeddings.

% Yunhao Tao
% Shiren Song
\textbf{(3) Hybrid Solution-Company Matching Framework (HSCM).} 
This is a previously online yet unpublished framework used by \huawei{}.
Based on the number of solutions in transaction data, the algorithm categorizes solutions into three groups. 
It devises unsupervised, semi-supervised, and supervised models for each category, respectively. 
It is a well-designed classification framework based on the classical Gradient Boosting Decision Tree (GBDT) model and BERT-encoded features.
Due to the company's confidentiality policy, it is not feasible for us to provide further elaboration. 
However, HCSM has the following drawbacks: (1) The framework maintains a distinct model for each solution, leading to a significant waste of resources. (2) Because of its naive structure, the results of the matching are unsatisfactory. 
% It is specifically poor at ranking companies, which can be observed by its relatively low AUC score.
% Shiren Song

\subsection{Implementation Details}

We use BERT provided by Huggingface as the token-level encoder\footnote{https://huggingface.co/hfl/chinese-bert-wwm-ext}.
The size of scale embedding $d_\text{s}$ is set as 64, and the parameters of AutoDis encoder are the same as its original paper~\cite{AutoDis}.
We use $k=6$ Transformer layers as the field-level encoder.
During the text matching enhancement process, we set the temperatures as 0.2 and 0.05 for pre-training description BERT ($\tau_d$) and attribute BERT ($\tau_a$), respectively.
The token mask ratio and field mask ratio are tuned and established as 0.2 and 0.5, respectively, for both BERTs.
% The analysis of these hyperparameters can be found in Section~\ref{subsec: hyper}.
% For deployment, we use a relatively lightweight yet effective LLM ChatGLM2-6B\footnote{https://github.com/THUDM/ChatGLM2-6B} for interpretation.
The learning rates are set as 3e-5 for the token-level encoding module, 5e-4 for the scale encoding module, and 5e-5 for both field-level encoding and pre-training.
The model is trained with a batch size of 32 for four epochs using four Tesla P100 16G GPUs.

\section{Results and Analysis}

To compare our proposed model \ours{} with baselines, we perform experiments in both offline and online settings. 
%However, due to the significant expenses of online marketing, 
We primarily conduct detailed experiments on BSM offline. 
We then deploy \ours{} on the online system to observe its overall performance.

\subsection{Overall Results} \label{subsec: overall results}

\subsubsection{Offline Results}

Experimental results on BSM are presented in Table~\ref{tab:result}.
The results show that the performance of \ours{} is significantly better than baseline models.
% We will notice that the two-tower SBERT performs better than single-tower BERT-Concat.
% We believe this is because that SBERT design 
We can make the following observations based on the results:

\textbf{(1) Our proposed method \ours{} outperforms all baselines.}
The offline results demonstrate that \ours{} performs significantly better than all baseline models.
% For example, our model outperforms a strong baseline HieGAT by about 17.4\% in terms of MAP on BSM.
% The results further validate the \textbf{non-trivial} nature of our problem.
This indicates that our hierarchical multi-field matching framework and contrastive pre-training technique are effective for matching solutions and companies.

\textbf{(2) Side-aware models generally perform better than the models purely based on text matching.}
For example, the weak side-aware model W\&D still outperforms the strong text-matching model CBERT.
This demonstrates the necessity of modeling the scale features.
Moreover, our model still performs better than all side-aware baselines, which demonstrates the effectiveness of our hierarchical multi-field matching framework again.

\subsubsection{Online Deployment Results}
To assess our proposed model, \ours{}, we implemented it in a real-world system over a six-month period. 
Our method, when given a solution slated for sale, generates a Top-K ranked list. 
The sales team for the solution then reviews this list, selecting companies based on the provided matching scores and their own expertise. 
Following this, salesmen contact the chosen companies and gather feedback to compute the Conversion Rate (CVR). 
Due to confidentiality constraints, we cannot reveal specific CVR figures. 
However, we present the relative CVR increase ratios during the evaluation period. 
Our approach involved having the same sales teams process both the company lists generated by our model and those produced by the pre-online model HSCM. 
By comparing the CVR results from these two models, \ours{} demonstrated a 29.99\% improvement over the HSCM model during the same timeframe.
% and a \textbf{52.85\%} improvement over the previously online model HSCM. 
% It's obvious that the metric of IP has a significant improvement over both manual recommendation and the previous online model HSCM. 
Moreover, we can make these observations:

\textbf{(1) Our proposed method \ours{} achieves a superior performance over HSCM.} The performance of \ours{} surpasses the previous online model HSCM by 29.99\%, which demonstrates its great commercial value.

\textbf{(2) Our proposed method \ours{} is not only better than the previous framework HSCM but also easy to deploy and maintain.} The online deployment results validate the effectiveness of the hierarchical multi-field matching structure and the contrastive pre-trained method compared with the improvement over HSCM. 
Furthermore, \ours{} provides a unified model for all solutions which makes it easy to maintain in online deployment.

% \begin{table}[t]
% \centering
% \caption{The overall performance of models in the online setting. ``Improvement'' indicates the improvement of each model over the manual recommendation method.} \label{tab:online}
% \begin{tabular}{cc}
% \toprule
% Model & Improvement \\
% \midrule
% Manual & - \\
% % Hybrid Solution-Company Matching Framework & 4 & 5 \\
% HSCM & 23\% \\
% \ours{} & 51\% \\
% \bottomrule
% \end{tabular}
% \end{table}
\begin{table}[t!]
    \centering
    \caption{Performances of ablated models on BSM.}\label{tab:ablation}
    \small
    \begin{tabular}{p{0.2\textwidth}cccc}
    \toprule
         Metric & \multicolumn{2}{c}{MAP} & \multicolumn{2}{c}{Rec@10}  \\
        \midrule
        \ours{} (Full) & \textbf{0.6810} & - & \textbf{0.4579} & - \\
        \quad w/o. Description Texts & 0.4891 & -28.18\% & 0.3905 & -14.72\%   \\
        \quad w/o. Attribute Texts & 0.5885 & -13.59\% & 0.4243 & -7.33\%   \\
        \quad w/o. Text Grouping & 0.4866 & -28.55\% & 0.3888 & -15.09\% \\
        \quad w/o. Field Embeddings & 0.5927 & -12.97\% & 0.4266 & -6.83\%  \\
        \quad w/o. Scale Encoding & 0.5596 & -17.83\% & 0.4162 & -9.11\%   \\
        \quad w/o. Field-level Interaction & 0.6386 & -6.23\% & 0.4393 & -4.07\% \\
        \quad w/o. Pre-training & 0.5919 & -13.09\% & 0.4259 & -6.98\%  \\
    \bottomrule%
    \end{tabular}%    
    \vspace{-4pt}
\end{table} %

\begin{figure*}[tbp]
\centering
\includegraphics[width=0.8\textwidth]{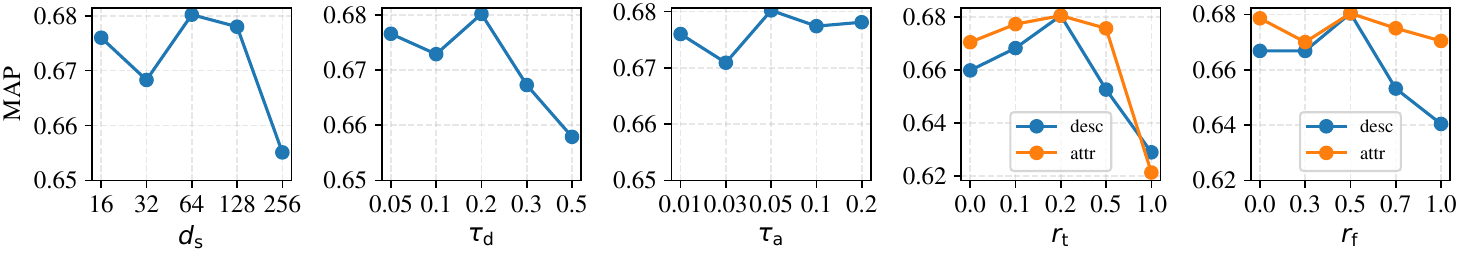}
\vspace{-2pt}
\caption{Performance of \ours{} on the BSM dataset with different hyperparameters.}\label{fig:hyper} 
\end{figure*}

\subsection{Ablation Studies} \label{subsec: ablation}

To evaluate the effectiveness of each component, we perform the following ablation studies on the BSM dataset: 

$\bullet$ \textbf{\ours{} w/o. Description Texts} is \ours{} without the interactions of Description Texts, \ie, without $P_{\text{desc}}$ and the corresponding field-level representations. 
$\bullet$ \textbf{\ours{} w/o. Attribute Texts} is \ours{} without the interactions of Attribute Texts.
$\bullet$ \textbf{\ours{} w/o. Text Grouping} is \ours{} without Text Grouping, \ie, we do not distinguish two types of texts and only compute one matching score ($ \{P_{\text{desc}}, P_{\text{attr}}\} \rightarrow P_{\text{text}}$ ).
$\bullet$ \textbf{\ours{} w/o. Field Embeddings} is \ours{} without the Field-aware Embeddings.
$\bullet$ \textbf{\ours{} w/o. Scale Encoding} is \ours{} without the Scale Encoding component.
$\bullet$ \textbf{\ours{} w/o. Field-level Interaction} is \ours{} without the Field-level Interaction module, \ie, $P_\text{field}$.
$\bullet$ \textbf{\ours{} w/o. Pre-training} is \ours{} without the pre-training of the token-level BERT encoders.

The results in Table~\ref{tab:ablation} clearly demonstrate that the full model outperforms all ablated models, indicating the effectiveness of our components. 
Moreover, we can make the following conclusions:

\textbf{(1) Both the interactions of two text groups can help the matching process.}
Specifically, \ours{}'s performance decreases by about 28.18\% and 13.59\% in terms of MAP after abandoning Description and Attribute Texts, respectively.
This indicates the importance of modeling both groups of texts.

\textbf{(2) The token-level interactions between two text groups interfere with each other.}
% We split them into two groups because they have different structures and meanings.
Our model's performance decreases by about 15.09\% in terms of Rec@10 after we combine two text groups and do not distinguish them in the token-level interaction module.
% This validates the necessity of modeling these two types of texts separately.

\textbf{(3) Identifying different text fields with field-aware embeddings is effective.}
The decrease of \ours{} after discarding Field-aware Embeddings demonstrates its effectiveness.

\textbf{(4) Modeling the scale of companies can facilitate our hierarchical textual interactions.}
% Specifically, the performance of \ours{} decreases by about 9.11\% in terms of Rec@10 after abandoning Scale Encoding.
This decrease of \ours{} after abandoning Scale Encoding meets our observation in Section~\ref{subsec: overall results}, that company scale plays an important role in modeling the matching between solutions and companies.
% However, the ablated model still performs better than all text matching baselines, which demonstrates the effectiveness of our text matching module.

\textbf{(5) Field-level interaction is effective for capturing inter-group matching signals.}
The performance of \ours{} decreases after abandoning the high-level interaction among fields of different groups.
This indicates the effectiveness of our field-level interaction.

\textbf{(6) Utilizing data augmentations and contrastive learning to pre-train our BERT encoders can make \ours{} more generalized.}
Specifically, our model's performance decreases by about 13.09\% in terms of MAP.
This validates that our data augmentation strategies and contrastive objective can make \ours{} more robust.

\subsection{Influence of Data Augmentation Strategies} \label{subsec: dataaug}

In our text matching enhancement module, we propose three data augmentation strategies and a contrastive learning objective to pre-train the BERT encoders.
% We believe this process can make our token-level encoders more generalized and robust with imperfect transaction data.
% Specifically, we design three data augmentation strategies to generate additional data pairs: Token masking, Field Masking, and Company Replacing.
To investigate the influence of these strategies, we compare \ours{} with the ablated models and show the results in Table~\ref{tab:data_aug}.
% For \ours{} w/o. Token Masking and \ours{} w/o. Field Masking, we will repeat the remaining masking strategy twice for the negative sequences since Company Replacing can only be done on positive pairs.
% The results demonstrate the effectiveness of our data augmentation strategies.
We can make the following observations:

\textbf{(1) Token masking strategy can help pre-train the BERT encoders.}
After abandoning the sequences generated by masking specific tokens, the performance of our model drops by about 4.86\% in terms of MAP.
This shows that the available transaction data are deficient for training a robust model, and our token masking strategy can help \ours{} get more generalized representations.

\textbf{(2) Pairs augmented by field masking can mitigate the problem of incomplete data.}
There are some solutions and companies lack specific data fields.
The performance of \ours{} w/o. Field Masking validates this problem of our data and demonstrates that our strategy can help mitigate this challenge.

\textbf{(3) Replacing the company in a matching pair with a similar company can generate data pairs for addressing data sparsity.}
% We believe that there are numerous unrecorded pairs in the transaction data due to potential subjective factors, which results in the data sparsity problem.
% We propose to mine a similar company to replace the original one in each transaction pair to address the data sparsity problem.
The performance of our model decreases by about 6.65\% in terms of Rec@10 after discarding Company Replacing, which demonstrates this technique can complement the sparse data.

\begin{table}[t!]
    \centering
    \caption{Performances of \ours{} without different data augmentation strategies on BSM.}
    \small
    \begin{tabular}{lcccc}
    \toprule
         Metric & \multicolumn{2}{c}{MAP} & \multicolumn{2}{c}{Rec@10}  \\
        \midrule
        \ours{} (Full) & \textbf{0.6810} & - & \textbf{0.4579} & - \\
         \quad w/o. Token Masking & 0.6479 & -4.86\% & 0.4463 & -2.54\%   \\
         \quad w/o. Field Masking & 0.6097 & -10.47\% & 0.4326 & -5.53\%   \\
         \quad w/o. Company Replacing & 0.5961 & -12.47\% & 0.4274 & -6.65\% \\
    \bottomrule
    \end{tabular}
    % \vspace{-5px}
    \label{tab:data_aug}
\end{table}

\subsection{Influence of Hyperparameters} \label{subsec: hyper}

\subsubsection{The Embedding Size of Company Scale}

% We propose to model the scale features to get an understanding of the company's financial situation and help the interactions of text fields.
To model the interactions between scale features and text embeddings, we have to embed the scale features into a vector $\mathbf{c}^\text{s} \in \mathbb{R}^{d_\text{s}}$.
We tune the scale embedding size $d_\text{s}$ in the range $[16, 256]$ with the step of 16 on the validation set and present the performances of \ours{} with different $d_\text{s}$ on the test set.
Due to the space limitation, we present the performance of MAP on BSM and only show the results of five tuned values (the henceforth displays will follow the same policy).
As shown in the left part of Fig.~\ref{fig:hyper}, the performances increase to the optimal value and then drop.
This pattern indicates a trade-off:
If the embedding size is too low, the scale embedding can not encode sufficient information.
However, it may introduce noise into \ours{} if the embedding size is too high.

\subsubsection{The Temperature of Contrastive Learning}

In our contrastive loss, there is a hyperparameter $\tau$ representing the temperature hyperparameter which controls the model’s discrimination against negative samples.
If it is set too low, our model will concentrate on the negative pairs that are hard to distinguish.
However, a high value of $\tau$ will make \ours{} treat all negative samples equally.
% We tune the temperatures for pre-training description BERT and attribute BERT separately.
We tune $\tau_d$ in the range $[0.05, 0.5]$ with the step of 0.05 and $\tau_a$ in the range $[0.01, 0.2]$ with the step of 0.01.
The results presented in the middle part of Fig.~\ref{fig:hyper} demonstrate the trade-off of this hyperparameter and validate our choices.

\subsubsection{The Mask Ratios}

In our data augmentation strategies, we use mask ratios $r_\text{t}$ and $r_\text{f}$ to control the number of tokens/fields we mask in Token Masking and Field Masking strategies, respectively.
We tune these two ratios for pre-training two BERTs and find that the patterns of the two BERTs are the same, thus we present the tuning results of each ratio for both BERTs in the same figure.
Specifically, we tune $r_\text{t}$ and $r_\text{f}$ in the range $[0.0, 1.0]$ with the step of 0.05.
The results shown in the right part of Fig.~\ref{fig:hyper} indicate there is also a trade-off for the masking ratios.
If we set the masking ratios too high, the augmented pairs may not be similar to the original pair.
However, too low masking ratios will introduce little knowledge into our pre-training process, resulting in insufficient pre-training.

\section{Conclusion}
% \section{Conclusion and Future Work}

In this work, we study a valuable yet understudied problem of B2B solution matching and identify two key challenges in this scenario.
Initially, we propose a hierarchical multi-field matching framework to model the interactions between the complex multi-field features of solutions and companies.
Subsequently, three data augmentation strategies and a contrastive learning objective are proposed to deal with the limited, incomplete, and sparse transaction data.
% Lastly, we extract valuable features and summarize them with a Large Language Model for interpretability.
Extensive experiments on a real-world dataset BSM demonstrate the effectiveness of \ours{}.
The deployment of our framework on \huawei{} validates the feasibility and effectiveness of our framework in an online scenario.
Considering the generalizability of our framework, it can also be applied to other B2B matching scenarios that encounter similar challenges.

\begin{acks}

Zhicheng Dou is the corresponding author. This work was supported by National Natural Science Foundation of China No. 62272467, the fund for building world-class universities (disciplines) of Renmin University of China, and Public Computing Cloud, Renmin University of China. The work was partially done at the Engineering Research Center of Next-Generation Intelligent Search and Recommendation, MOE, and Beijing Key Laboratory of Big Data Management and Analysis Methods.

\end{acks}

% In the future, \todo{}.

% \section*{Acknowledgment}

% \clearpage
% \bibliographystyle{IEEEtran}

% \bibliographystyle{ACM-Reference-Format}
% \bibliography{references}
\input{main.bbl}

\end{document}